# Segmentasi Citra Menggunakan Metode Watershed Transform Berdasarkan Image Enhancement Dalam Mendeteksi Embrio Telur

**Shoffan Saifullah**

Universitas Pembangunan Nasional Veteran Yogyakarta

shoffans@upnyk.ac.id

| *Kata Kunci* | *Abstrak* |
|---|---|
| *Image Processing, Deteksi Embrio Telur, CLAHE-HE, Watershed Transform* | *Image processing dapat diterapkan dalam proses deteksi embrio telur. Proses deteksi embrio telur dilakukan dengan menggunakan proses segmentasi, yang membagi citra sesuai dengan daerah yang dibagi. Proses ini memerlukan perbaikan citra yang diproses untuk memperoleh hasil optimal. Penelitian ini akan menganalisis deteksi embrio telur berdasarkan image processing dengan image enhancement dan konsep segmentasi menggunakan metode watershed transform. Image enhacement pada preprocessing dalam perbaikan citra menggunakan kombinasi metode Contrast Limited Adaptive Histogram Equalization (CLAHE) dan Histogram Equalization (HE). Citra grayscale telur diperbaiki dengan menggunakan metode CLAHE, dan hasilnya diproses kembali dengan menggunakan HE. Hasil perbaikan citra menunjukkan bahwa metode kombinasi CLAHE-HE memberikan gambar secara jelas daerah objek citra telur yang memiliki embrio. Proses segmentasi dengan menggunakan konversi citra ke citra hitam putih dan segmentasi watershed mampu menunjukkan secara jelas objek telur ayam yang memiliki embrio. Hasil segmentasi mampu membagi daerah telur memiliki embrio secara nyata dan akurat dengan persentase sebesar $\approx 98\%$..* |
| *Keywords* | *Abstract* |
| *Image Processing, Detection of Egg Embryos, CLAHE-HE, Watershed Transform* | *Image processing can be applied in the detection of egg embryos. The egg embryos detection is processed using a segmentation process. The segmentation divides the image according to the area that is divided. This process requires improvement of the image that is processed to obtain optimal results. This study will analyze the detection of egg embryos based on image processing with image enhancement and the concept of segmentation using the watershed method. Image enhancement in preprocessing in image improvement uses a combination of Contrast Limited Adaptive Histogram Equalization (CLAHE) and Histogram Equalization (HE) methods. The grayscale egg image is corrected using the CLAHE method, and the results are reprocessed using HE. The image improvement results show that the CLAHE-HE combination method gives a clear picture of the object area of the egg image that has an embryo. The segmentation process using image conversion to black and white image and watershed segmentation can clearly show the object of a chicken egg that has an embryo. The results of segmentation can divide the area of the egg having embryos in a real and accurate way with a percentage $\approx 98\%$.* |

## 1. Pendahuluan

Deteksi objek dapat dilakukan dalam beberapa cara diantaranya dengan suara [1], [2], gambar [3], dan sensor [4]. Segmentasi merupakan salah satu proses yang digunakan untuk mendeteksi objek gambar. Segmentasi mampu membagi citra [5] ke dalam beberapa bagian (*region*). Proses segmentasi citra dilakukan setelah *preprocessing*. *Preprocessing* dilakukan untuk mendapat gambar yang lebih baik secara kualitas untuk dilakukan proses identifikasi. Beberapa proses yang dilakukan untuk memperbaiki citra diantaranya *image enhacement* menggunakan metode *contrast limited adaptive histogram equalization* (CLAHE) dan *histogram thresholding* (HT). Image enhacement dengan metode CLAHE dan HT mampu memperbaiki citra objek bawah air [6] sehingga hasil yang diberikan menunjukkan gambar dengan objek yang lebih jelas dibandingkan dengan citra aslinya.

Selain dengan menggunakan CLAHE, *image enhancement* dapat dilakukan dengan menggunakan metode *Histogram Equalization* (HE) dan *Exposure Based Sub Image Histogram Equalization* (ESIHE) [7]. Perbandingan hasil diperoleh bahwa ESIHE mampu memberikan hasil yang lebih baik jika dibandingkan dengan metode HE dan CLAHE dalam peningkatan kualitas berdasar pada nilai *entropy* dan *contrast* [7], [8], [9].

Peningkatan kualitas pada citra *myocardial perfussion*, metode CLAHE mampu memberikan hasil yang efektif dan cepat dalam interpretasi visual [9], [10] dan lebih akurat dalam analisis diagnosa pada *cardiac*, jika dibandingkan dengan metode HE. Metode HE memberikan *intersity saturation* yang hasilnya menginformasikan bahwa metode ini belum dapat diimplementasikan pada kasus citra medis [11]. Hasil CLAHE menggambarkan citra yang lebih jelas dan dapat dilakukan proses segmentasi seperti penggunaan metode *winner filter* untuk proses segmentasi citra biomedical. Selain dengan metode-metode tersebut, *image enhacement* dapat dilakukan dengan menggunakan beberapa metode seperti simpel HE [12], *Dualistic Sub-Image Histogram Equalization* (DSIHE) dan HE *Schenaus* [13].

Selain proses perbaikan citra (*image enhancement*), proses segmentasi sangat penting dalam proses identifikasi [14]. Proses segmentasi telur telah dilakukan dalam identifikasi citra asli dan citra hasil *watermarking* [15]. Hasil segmentasi menunjukkan bahwa prosesnya berhasil dan dapat terindentifikasi. Proses segmentasi dilakukan dengan berbagai metode diantaranya yaitu *thresholding* [16], HSV (*hue, saturation, value*) [17], *active contour* [18] [19], *otsu method* [20], *edge detection* [21], *hough transformation* [22], watershed [23], dan lain-lain.

Proses identifikasi citra telur diproses dengan berbagai cara mulai dari pengenalan fisik maupun embrio telur [24]. Proses identifikasi telah dilakukan dalam beberapa penelitian diantaranya yaitu identifikasi fertilitas telur ayam berdasarkan ektrasksi ciri GLCM dengan Backpropagatoin dan K-Means Clustering [25]. Selain itu, proses identifikasi citra telur hasil *thermal imaging* [26] dapat diproses dengan *image processing*. Identifikasi citra juga diproses untuk menguji image processing dalam perbandingan proses diantaranya yaitu citra dan hasil kompresi wavelet-nya [27], citra asli dan hasil *croping*-nya [28], citra digital dan thermal imaging [29], proses segmentasi citra asli dengan citra hasil *watermarking* [30], ekstraksi fitur GLCM [31] dan Backpropagation (BP) [32] dalam identifikasi fertilitas telur.

Berdasarkan pada latar belakang tersebut, maka dalam penelitian ini akan mendeteksi gambar embrio telur dengan menggunakan *image enhacement* (kombinasi CLAHE dan HE), dan proses segmentasi dengan metode watershed.

## 2. Metode Penelitian

Identifikasi telur dengan menggunakan *image processing* banyak dilakukan dengan berbagai metode dan berbagai hasil identifikasi. *Image processing* merupakan salah satu proses identifikasi fertilitas telur dilakukan dengan langkah awal akuisisi citra. Proses akuisisi citra dapat dilakukan dengan beberapa alat yaitu kamera digital dan kamera thermal [15], [28] [33]. Proses identifikasi berdasarkan image processing salah satunya menggunakan metode segmentasi untuk menentukan objek yang teridentifikasi sebagai telur.

Proses identifikasi telur yang dilakukan adalah mendeteksi embrio pada telur menggunakan metode segmentasi watershed. Citra telur yang digunakan adalah citra telur ayam. Citra telur akan dilakukan *preprocessing* terlebih dahulu dengan guna untuk memperbaiki citra sehingga hasil prosesnya dapat maksimal. Proses deteksi embrio telur ditunjukkan seperti pada Gambar 1.

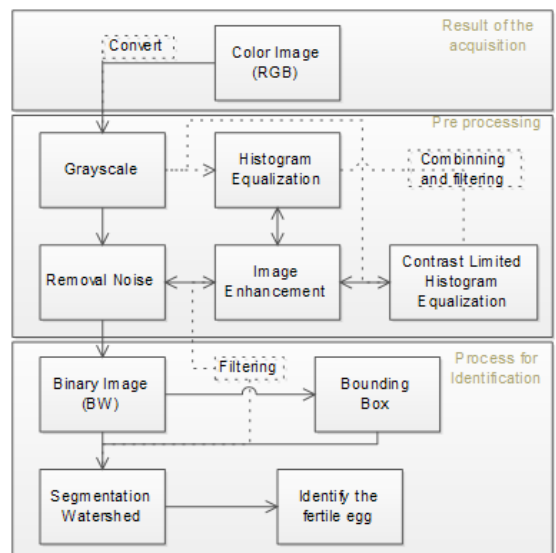

*Gambar 1. Alur proses deteksi embrio pada telur fertil dengan menggunakan segmentasi watershed*

Berdasarkan pada Gambar 1, alur proses memiliki 3 tahap utama yaitu hasil akuisisi citra, *preprocessing*, dan identifikasi. Proses akuisisi citra menghasilkan citra warna (RGB) dengan 3 komponen warna yaitu (*red*), hijau (*green*), biru (*blue*) [34], [35]. Hasil ini akan dilakukan rposes *grayscaling*, yaitu mengubah citra warna menjadi citra *grayscale*. Dimana citra warna direpresentasikan ke dalam matriks m x n dengan 3 komponen warna tersebut yang masing-masing warna memiliki range 0-255. Proses grayscaling ini dilakukan dengan menggunakan rumus persamaan (1).

$$grayscale = 0,299R + 0,587G + 0,114B \quad (1)$$

Citra *grayscale* yang dihasilkan perlu dilakukan perbaikan citra dengan menghilangkan *noise*/derau yang ada pada citra. Noise merupakan gangguan pada citra yang dapat menurunkan kualitas citra [36].

*Image enhancement* merupakan proses peningkatan kualitas yang bertujuan untuk memperbaiki citra maupun menunjukkan citra yang ditonjolkan [37]. Perbaikan citra pada penelitian ini menggunakan kombinasi 2 metode yaitu *histogram equalization* dan *contrast limited adaptive histogram equalization* (CLAHE).

HE merupakan metode yang digunakan untuk merenggangkan/meratakan histogram [38]. Hasil HE memberikan perbedaan piksel yang lebih besar sehingga informasi yang ada dari citra mampu ditangkap oleh mata. HE akan memperlebar range tingkat keabuan dari citra *grayscale* untuk meningkatkan konstras citra. HE dapat diperoleh dengan memproses nilai keabuan citra dengan rumus persamaan (2).

$$s_k = T(r_k) = \sum_{j=0}^{k} \frac{n_j}{n} = \sum_{j=0}^{k} p(r_j), \ 0 \le r_k \le 1, \ k = 0,1,2,..,L-1 \quad (2)$$

dimana *nk* adalah nilai piksel derajat keabuan, *k* dan *n* merupakan jumlah piksel dan *L* adalah derajat keabuan. Sehingga nilai derajat keabuan (*k*) dinormalkan terhadap derajat keabuan (*L-1*). Jika nilai *rk* = 0 maka warna hitam, dan jika *rk*= 1 maka warna putih (dalam skala keabuan).

CLAHE memiliki konsep dimana histogram yang diproses diberi nilai batas bawah [39]. CLAHE menunjukkan kejelasan pada citra dan konstras citra tidak terlalu meningkat. CLAHE dapat mencegah noise over enhacement dan mengurangi *edge shadowing* [40]. Nilai batas ini disebut dengan *clip limit* yang dapat dihitung dengan rumus persamaan 3.

$$\beta = \frac{M}{N}\left(1 + \frac{a}{100}(s_{max} - 1)\right) \quad (3)$$

dimana M adalah luas *region size*, N merupakan nilai grayscale dengan range 256, dan *a* adalah clip factor sebagai batasan limit histogram antara 0-100.

Proses terakhir adalah identifikasi dengan mengubah citra hasil *preprocessing* menjadi citra hitam putih (BW). Citra BW difilter dengan menggunakan bounding box dan dilakukan proses segmentasi untuk identifikasi yang menunjukkan embrio telur. Pada bagian ini segmentasi menjadi metode akhir untuk proses deteksi embrio. Proses segmentasi dilakukan menggunakan *watershed transform*.

Konsep segmentasi didasari 2 pendekatan utama, yaitu tepi dan wilayah. Pendekatan segmentasi didasarkan pada tepi, pembagian citra dilakukan berdasarkan diskontinuitas antara sub wilayah. Sedangkan pendekatan segmentasi didasarkan pada wilayah, pembagian citra dilakukan berdasarpada keseragaman sub wilayah. Hasil segmentasi berupa bagian-bagian dari wilayang yang berkumpul dan melingkupi citra [41]. Segmentasi memiliki beberapa metode salah satunya adalah *watershed transformation* (disebut watershed). Konsep dasar watershed adalah menangkap citra dalam bentuk 3 dimensi (x,y,z dengan masing-masing tingkat warna piksel). Posisi x dan y adalah bidang dasar sebagai lokasi piksel, sedangkan z merupakan tingkat warna piksel grayscale dengan nilai yang semakin terang (mendekati putih) semakin tinggi nilai ketingiannya. Proses segmentasi watershed dapat dilakukan dengan menggunakan teknik *flooding*. Teknik *flooding* dilakukan dengan langkah-angkah sebagai berikut:

a. Proses penentuan jarak tranformasi dengan menggunakan Euclidean Distance (ED) berdasarkan citra hasil *preprocessing*. Rumus ED dapat ditunjukkan pada persamaan 4.

$$D = \sqrt{(x_1 - y_1)^2 + (x_2 - y_2)^2} \quad (4)$$

b. Mencari nilai jarak terkecil (persamaan 5)

$$d = \min(D_1, D_2, D_3, ..., D_n) \quad (5)$$

c. Membuat array yang digunakan untuk menerima setiap hasil nilai jarak dan label dari watershed dengan ukuran yang sama dengan citra input.

d. Mengurutkan nilai jarak secara *ascending*, dan membuat variable inisiasi dan *increment* untuk proses *flooding*.

e. Iterasi nilai-i, dimana untuk nilai setiap piksel yang sama dengan variable i, akan dicek terhadap 8 arah piksel tetangganya. Jika piksel tidak saling bersinggungan atau tidak saling bertumpuk, maka akan diberi label dan ditandai sebagai objek. Akan tetapi jika sebaliknya, maka piksel akan diberi label sebagai garis watershed dan disimpan pada array matriks label.

f. Mengulangi langkah ke-6 dengan nilai *increment*-nya ditambah dengan nilai terkecil dan jarak transformasi (dimana *i=i+d$_n$*)

Hasil akhir dilakukan pengecekan tingkat keakurasian [43] dengan rumus persamaan 6.

$$akurasi = \frac{TP + TN}{TP + TN + FP + FN} \quad (6)$$

dimana nilai dari semua komponen yang berhasil (*true positive*/TN dan *true negative*/TN) dibagi dengan seluruh data pengujian (semua komponen yang berhasil dan yang tidak (*false positive*/FP dan *false negative*/FN)) [44].

## 3. Hasil Dan Pembahasan

Proses identifikasi telur untuk mendeteksi embrio dilakukan dengan menggunakan citra telur ayam yang sudah diketahui bahwa telur tersebut fertil. Proses awal dalam pengolahan citra (akuisisi) dilakukan dengan menggunakan kamera *smartphone*. Kamera smartphone digunakan untuk akuisisi pengambilan gambar telur yang dilakukan proses *candling* (penyinaran telur dengan menggunakan senter) di tempat yang gelap seperti pada Gambar 2.

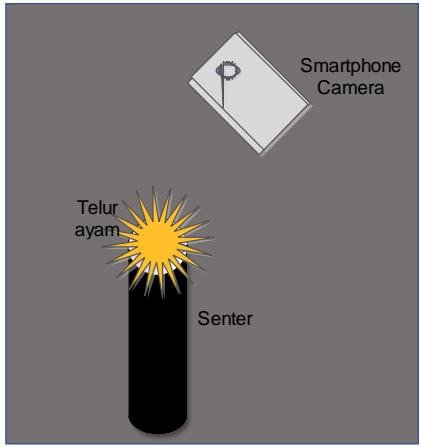

*Gambar 2. Rancangan Proses Akuisisi dari Proses Candling Citra Telur di Tempat Gelap dengan menggunakan senter dan kamera smartphone*

Akusisi citra (Gambar 2) akan menghasilkan citra warna (RGB) seperti pada Gambar 3.(a). Hasil proses akusisi citra (citra warna) akan dilakukan *preprocessing*, dimana langkah awalnya adalah dengan proses *grayscaling* (hasilnya seperti pada Gambar 3.b).

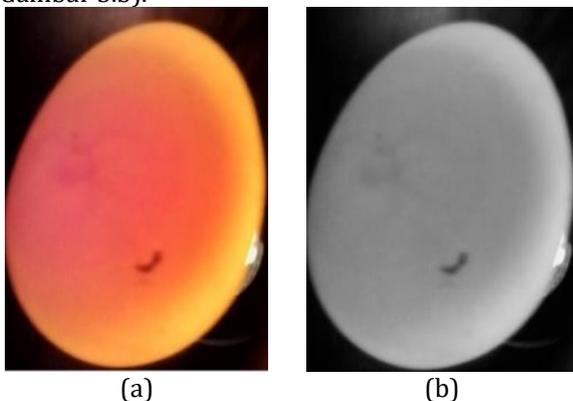

*Gambar 3. Gambar Telur Ayam Fertil (a) Gambar Telur Warna, (b) Gambar Telur Grayscale*

Gambar 3 menunjukkan data citra telur yang terdeteksi sebagai telur yang fertil yang di dalamnya terdapat embrio. Proses identifikasi awal adalah dengan menggubah citra telur warna (Gambar 3.(a)) menjadi citra *grayscale* (Gambar 3.b). Proses ini dilakukan dengan menggunakan nilai matriks dari citra warna dengan masing-masing komponen warna (R,G,dan B) dimasukkan ke dalam persamaan 1 akan menghasilkan citra *grayscale* (citra aras keabuan dengan intensitas nilai dengan range 0-255)

Hasil *grayscaling* (*citra grayscale*) secara detil belum memberikan gambar yang jelas adanya embrio dalam telur, akan tetapi ketika dilihat lebih fokus dan detail maka dibagian dalam sudah tergambarkan adanya embrio yang menandakan bahwa telur tersebut dikategorikan dalam telur fertil. Setelah proses *grayscale* dilakukan *preprocessing* menggunakan metode HE dan CLAHE dengan menggunakan masing-masing dan gabungan dari kedua metode tersebut (Gambar 4).

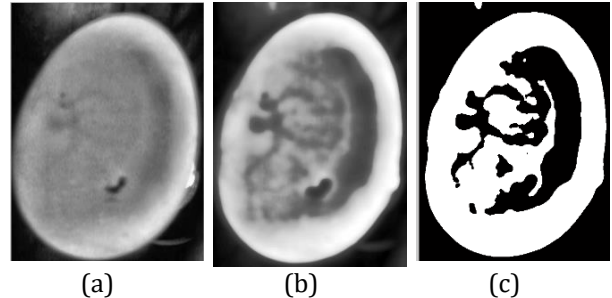

(a) (b) (c)
*Gambar 4. Hasil Pengolahan Citra (a) Proses CLAHE (b) Proses CLAHE dan HE, (c) Black White*

Proses pengolahan citra dengan menggunakan metode CLAHE (Gambar 4.(a)) memberikan gambar dengan warna kontras yang berada di dalam lebih terlihat jelas jika dibandingkan dengan citra hasil *grayscale*, akan tetapi masih memiliki kekurangan dengan banyaknya *noise*. Karena hal tersebut, citra hasil CLAHE diproses menggunakan metdeo HE yang menggambarkan hasil lebih jelas (Gambar 4.(b)). Hasil CLAHE-HE diproses dengan removal noise, konversi ke citra BW, dan *filtering* sehingga menghasilkan citra BW yang sangat jelas menggambarkan adanya embrio di dalam citra telur tersebut (Gambar 4.(c)).

Perbandingan p*reprocessing* dengan *image enhancement* metode yang digunakan adalah HE dan CLAHE, dimana dilakukan analisis histogram. Histogram yang dianalisis yaitu citra hasil penerapan metode CLAHE, HE, kombinasi HE-CLAHE, dan CLAHE-HE. Masing-masing metode akan ditampilkan gambar histogramnya.

Perbandingan penerapan metode ditunjukkan pada Gambar 5. Penerapan dengan metode CLAHE ditunjukkan seperti gambar 5.(a) dimana sebelah kiri menggambarkan hasil citranya sedangkan sebelahnya menunjukkan histrogram yang dihasilkan. Gambar masih terdapat banyak *noise* dan gambar kurang jelas. Histogram yang dihasilkan menunjukkan gambar yang masih terkumpul. Sedangkan gambar 5.(b) menunjukkan hasil HE yang histogramnya sudah menyebar rata dan merenggang, dan gambar yang dihasilkan masih kurang jelas.

Hasil HE-CLAHE dan CLAHE-HE memberikan perbedaan bahwa citra CLAHE-HE (Gambar 5.(c)) memberikan gambar yang lebih jelas dibandingkan dengan HE-CLAHE (dari Gambar 5.(d)). Sehingga hasil yang diproses lanjut adalah Gambar 5.(d). Gambar 5.(d) merupakan hasil akhir *preprocessing* yang akan diproses segmentasi.

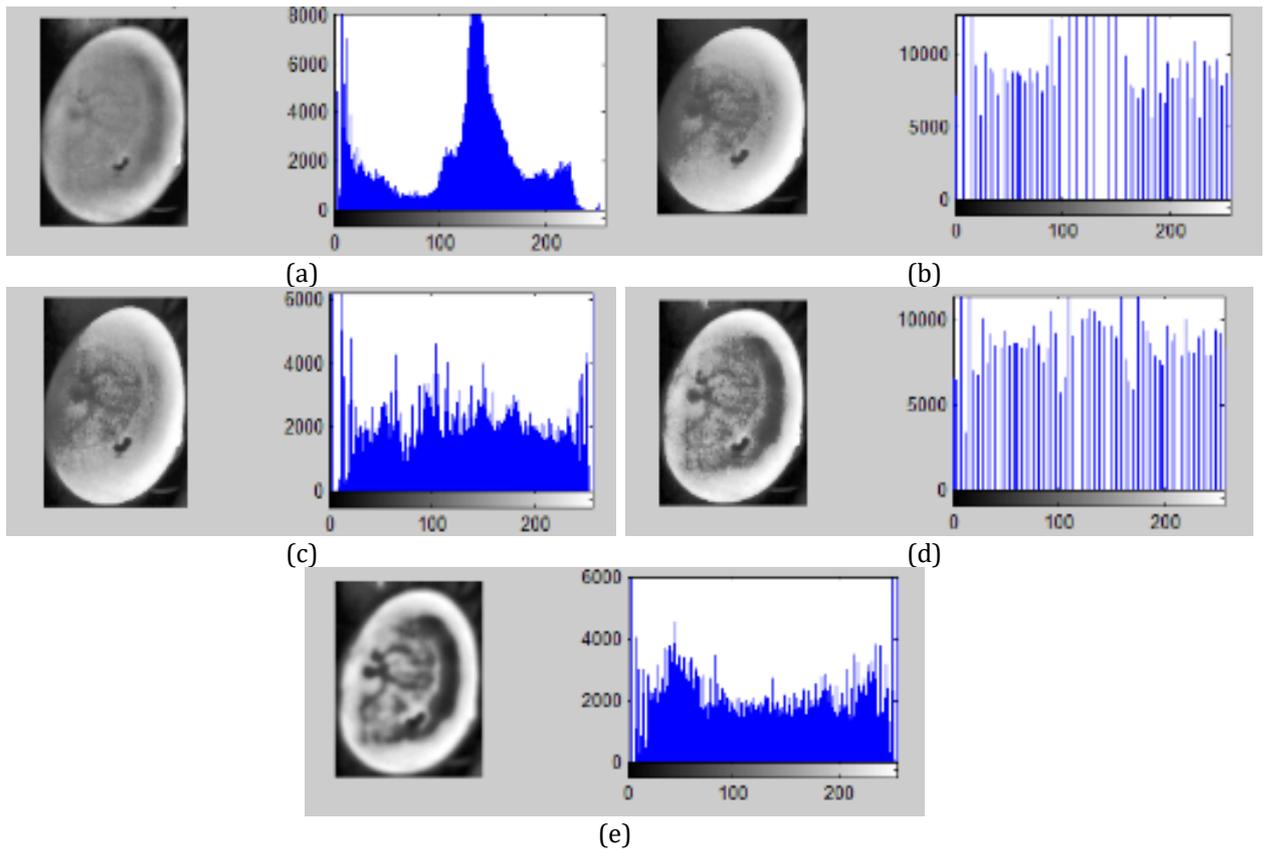

*Gambar 5. Analisis Metode Histogram Equalization dengan metode (a) CLAHE, (b) HE, (c) HE-CLAHE, (d) CLAHE-HE, dan (e) hasil removal noise dan filtering dari CLAHE-HE*

Segmentasi menggunakan metode watershed yang ditunjukkan seperti pada Gambar 6. Proses dilakukan mulai dari *removal noise* dan *filtering* dengan *bounding box* yang ditunjukkan dari Gambar 6.(a)-6.(c). Kemudian akan dipeoses pembuatan citra BW dengan hasil negasinya Gambar 6.(d)-6.(e). Proses terakhir adalah menggunakan segmentasi *watershed* yang ditunjukkan Gambar 6.(f)-6(h).

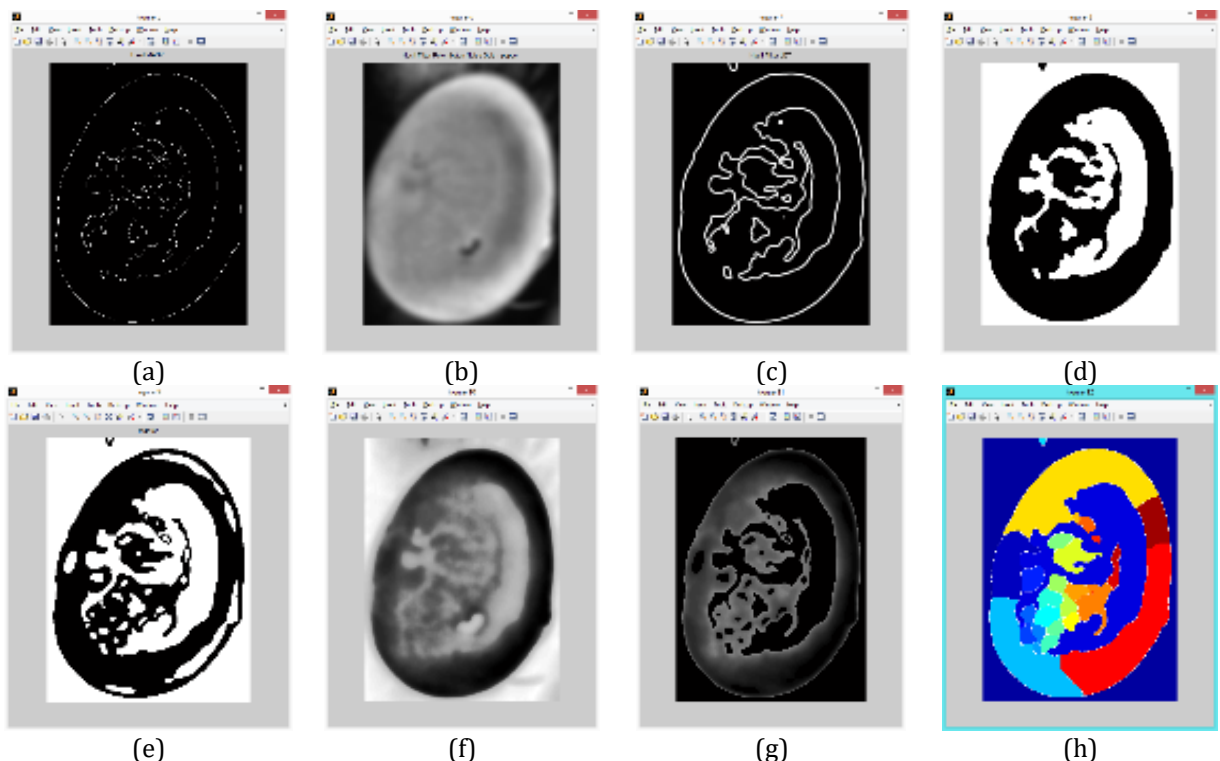

*Gambar 6. Analisis Proses Watershed dari awal sampai akhir proses untuk masing-masing citra hasil.*

Hasil segmentasi watershed, hasilnya ditunjukkan seperti pada Gambar 6.(h) dimana pada bagian akhir citra akan dilabeli dengan warna. Citra telur yang memiliki embrio terlihat jelas dengan warna-warna yang ditunjukkan dan menjadi dasar jika terjadi banyak segmentasi dengan *watershed* akan menunjukkan secara jelas embrio yang ada di dalam telur. Berdasarkan proses deteksi embrio, keakurasian hasil deteksi menunjukkan persentase sekisar 98%. Hal ini dapat digunakan untuk memilah telur yang memiliki embrio dan tidak pada penetasan telur. Jika telur terdeteksi memiliki embrio maka tetap dilakukan penetasan, jika tidak memiliki embrio akan dikeluarkan dari penetasan.

## 4. Kesimpulan

Deteksi embrio pada citra telur menggunakan metode *watershed transform* dilakukan dengan membagi daerah/bagian objek pada telur dengan warna yang berbeda. Hasil deteksi menunjukkan bahwa gambar isi telur memiliki embrio. Embrio yang terdeteksi terlihat jelas dalam telur, ditandai dengan adanya garis-garis seperti akar di dalam telur ayam dan terdapat bulatan kuning yang telihat secara jelas. Akan tetapi citra hasil akuisisi belum mampu memberikan deteksi pada citra telur, sehingga perlu dilakukan proses *image enhancement* dengan menggabungkan 2 metode yaitu CLAHE dan HE sehingga mampu memberikan gambar yang jelas dan mudah untuk dilakukan proses segmentasi dengan menggunakan metode *watershed transform*. Sehingga hasil yang didapatkan mampu menunjukkan bahwa metode *image enhancement* dan *watershed transform* dapat digunakan dalam proses identifikasi pada citra telur berembrio dengan persentase keberhasilan sekisar 98%.